\documentstyle[12pt]{article}
\def\e{\epsilon}
\def\A{{\cal A}}
\def\B{{\cal B}}
\def\M{{\cal M}}
\def\S{{\cal S}}
\def\T{{\cal T}}

\begin{document}

\begin{titlepage}
\vspace*{-1cm}
\begin{flushright}
hep-ph/9710255\\
DTP/97/82\\
\end{flushright}
\vskip 1.cm
\begin{center}
{\Large\bf
Double Unresolved Approximations to Multiparton Scattering Amplitudes}
\vskip 1.3cm
{\large J.~M.~Campbell   and  E.~W.~N.~Glover}\\
\vspace{0.5cm}
{\it
Physics Department, University of Durham,\\ Durham DH1~3LE, England} \\
\vspace{0.5cm}
{\large \today}
\vspace{0.5cm}
\end{center}

\begin{abstract}
We present approximations to tree level multiparton scattering amplitudes
which are appropriate when two partons are unresolved.
These approximations are required for the analytic isolation of infrared
singularities
of $n+2$ parton scattering processes contributing to the
next-to-next-to-leading
order corrections to $n$~jet cross sections.
In each case the colour ordered matrix elements factorise and yield a function
containing the singular factors multiplying the $n$ parton amplitudes.
When the unresolved particles are colour unconnected, the approximations
are simple products of the familar eikonal and
Altarelli-Parisi splitting functions used to describe single unresolved
emission.
However, when the unresolved particles are colour connected the factorisation
is more complicated and we introduce new and general functions to describe the
triple
collinear and soft/collinear limits in addition to the known double soft
gluon limits of Berends and Giele.
As expected the triple collinear splitting functions obey an $N=1$ SUSY
identity.
To illustrate the use of these double unresolved approximations, we have
examined the singular limits of the tree level matrix elements for
$e^+e^- \to 5$~partons when only three partons are resolved.
When integrated over the unresolved regions of phase space, these expressions
will be of use in evaluating the ${\cal O}(\alpha_s^3)$
corrections to the three jet rate in electron positron annihilation.

\end{abstract}
\vfill
\end{titlepage}
\section{Introduction}
\setcounter{equation}{0}
\label{sec:int}

Electron-positron annihilation has proved a very clean and
direct source of information about the nature of the strong interaction.
In particular, three jet events observed at DESY gave the first clear
indications
of the existence of the gluon \cite{gluon}, while more recently,
the Casimirs that determine the gauge group of QCD have been
measured \cite{casimir} using four jet data.
On the theoretical side,
a large amount of effort has been devoted to
determining multi-jet rates within perturbative QCD.
Leading order predictions for the production of up to five jets
have been known for some time \cite{egr,ali,ert,fsks,5jet}.
While the general features of events are generally well described by
leading order
estimates, significant improvement can be obtained by including higher
order corrections at either fixed order or via resummation of the dominant
logarithms or a mixture of both.
The next-to-leading order corrections to three-jet like quantities
were first computed in the early 1980's \cite{ert,fsks} and systematically
evaluated by Kunszt and Nason \cite{KN} using a general purpose
Monte Carlo program.
In many cases the radiative corrections are significant and
resummations of infrared logarithms \cite{resum}
must be employed to
make sensible comparisons with experimental data.
Such calculations have been used to extract a precise value
of the strong coupling constant from three jet events and other hadronic
event shapes \cite{schmelling} with a global average of,
$$
\alpha_s(M_Z) = 0.121 \pm 0.005.
$$
A recent re-evaluation of lower energy data yields \cite{siggi},
$$
\alpha_s(M_Z) = 0.122 ^{+0.008}_{- 0.006} ,
$$
while preliminary analyses using hadronic data above the $Z$-resonance
from the OPAL Collaboration \cite{opal} shows that the strong coupling
evolves in the manner predicted by QCD.

Recently there has been much progress towards a more complete
theoretical description of four jet events produced in
electron-positron annihilation.
The one-loop matrix elements relevant for four jet production
have been evaluated by two groups
\cite{GM,BDKW,CGM,BDKall}
using contrasting methods.
Bern et al calculate helicity amplitudes for the
$e^+e^- \to (\gamma^*,Z) \to 4$~parton processes
while in Refs.~\cite{GM,CGM}, the spin summed
interferences between one-loop and tree-level $\gamma^* \to 4$~parton
processes were evaluated.   These results have been compared \cite{BDKall} and
found to agree for specific phase space points.
Using these one-loop matrix elements and the earlier tree level five parton
matrix elements of \cite{5jet}, two general purpose Monte Carlo programs
have been constructed  \cite{DS,NT} and used to compute
the four-jet rate and other four-jet like hadronic observables at
next-to-leading order (i.e. ${\cal O}(\alpha_s^3)$).
The corrections are again large \cite{DS,NT} indicating the need
for resummations of
infrared logarithms, but nevertheless, improved determinations of
the Casimirs are possible \cite{NTcasimir} together with more
stringent bounds on the existence of a light
supersymmetric partner of the gluon \cite{NTgluino}.

One of the next steps is to evaluate the next-to-next-to-leading order
corrections to the three-jet rate.
Together with LEP/SLC data and hadronic events from the next
linear collider
this would play a role in reducing the error on $\alpha_s(M_Z)$
from event shape analyses to the 2-3\% level.
However, to accomplish this, several ingredients are necessary.
First, the two-loop $\gamma^* \to 3$~parton amplitudes must be
determined including the evaluation of the two loop box graph with
massless internal and
external legs.
Second, the real infrared singularities which occur when
two particles are unresolved in the tree level process and when a
single
particle is unresolved in the one-loop process must be systematically
isolated.
In the case of two real unresolved particles, there are a variety
of different configurations;
\begin{itemize}
\item[{(a)}]  two soft particles,
\item[{(b)}]  two pairs of collinear particles,
\item[{(c)}]  three collinear particles,
\item[{(d)}]  one soft and two collinear,
\end{itemize}
while in the one-loop process one particle may be soft or two may be
collinear as before.
For next-to-leading order calculations,
many methods have been developed to first isolate the infrared singularities
associated with one unresolved particle
analytically and then numerically combine the remaining finite real
and virtual contributions \cite{fsks,slice,ert,subtract,hybrid}.
All of these rely on suitable approximations to the real matrix elements
in the soft and collinear limits.
However, the infrared singular limits of the real and one-loop amplitudes
relevant at next-to-next-to-leading order are rather less well known.
In ref.~\cite{loopcol}, Bern et al., have developed appropriate
splitting functions for one-loop processes where two external particles
are collinear while
Berends and Giele have examined the multiple soft gluon behaviour of QCD
matrix elements in ref.~\cite{multsoft}.

In this paper, and as a first step towards the next-to-next-to-leading
order 3 jet rate, we examine the double unresolved limits of multiparton
scattering amplitudes, and find suitable approximations for the
$e^+e^- \to 5$~parton matrix elements \cite{5jet}
when two particles are unresolved.
We present results for all of the configurations (a)-(d) above.
In section~\ref{sec:single}, we briefly review the structure of tree level
scattering amplitudes when one particle is unresolved.
In section~\ref{sec:nnlo} we
write down general expressions for the structure of the $e^+e^- \to 5$~
parton matrix elements.
Sections~\ref{sec:unconnected} and \ref{sec:connected} are organised
according to whether the two unresolved particles are colour connected or not.
The precise meaning of what colour connected means will be given in
sect.~\ref{sec:unconnected}.
In the unconnected case, the singular limits are merely obtained by
multiplying single unresolved factors.   However,
when the particles are colour connected, the structure is more
involved (sect.~\ref{sec:connected})
and we give explicit formulae detailing the double unresolved singular factors.
In each case we write down expressions for the double unresolved limits of the
five parton process.
Our results are summarised in section~\ref{sec:conc}.

\section{Single unresolved limits}
\setcounter{equation}{0}
\label{sec:single}

The soft gluon and collinear parton limits of multiparton
scattering amplitudes are well known. For example, if we
consider the matrix element for the tree level
scattering of $n$ gluons labelled $1\ldots n$, then,
decomposing the amplitude according to the various colour
structures \cite{coldec}, we find,
\begin{equation}
{\cal M}(1,2,\ldots,n) = 2 i g^{n-2} \sum_{P(1,2,\ldots,n-1)}
Tr(T^{a_1}T^{a_2}\ldots T^{a_n}) \A(1,2,\ldots,n).
\end{equation}
Here, $a_i$ represents the colour of the $i$th gluon and
the summation runs over the $(n-1)!$ non-cyclic permutations of the gluons.
At leading order in the number of colours, $N$,
the squared matrix elements summed over all colours have the simple form,
\begin{equation}
|{\cal M}(1,2,\ldots,n)|^2 = \left(\frac{g^2N}{2}\right)^{n-2} (N^2-1)
   \sum_{P(1,2,\ldots,n-1)} |\A(1,2,\ldots,n)|^2.
\end{equation}
Unlike QED \cite{YFS}, it is the colour ordered sub-amplitudes
${\cal A}$ rather than the squared matrix elements that have nice
factorisation properties in the soft and collinear limits
(see for example~\cite{current}).
For example, in the limit where the $n$th gluon becomes soft,
we have the QED-like factorisation into an eikonal factor
multiplied by the colour ordered
amplitude with gluon $n$ removed, but the ordering of the hard gluons
preserved,
\begin{equation}
|\A(1,2,\ldots,n-1,n)|^2 \to S_{(n-1)n1}(s_{1(n-1)},s_{1n},s_{n(n-1)})\,
|\A(1,2,\ldots,n-1)|^2.
\end{equation}
with the eikonal factor given by,
\begin{equation}
S_{abc}(s_{ac},s_{ab},s_{bc})=\frac{4s_{ac}}{s_{ab}s_{bc}}.
\label{eq:ssoft}
\end{equation}
Similarly, in the limit where two gluons become collinear, the sub-amplitudes
factorise.
For example, if gluons $a$ and $b$ become collinear and form
gluon $c$,  then  adjacent gluons give a singular contribution,
\begin{equation}
|\A(1,\ldots,a,b,\ldots,n)|^2 \to P_{gg\to g}(z,s_{ab})\,
|\A(1,\ldots,c,\ldots,n)|^2,
\end{equation}
while separated gluons do not,
\begin{equation}
|\A(1,\ldots,a,\ldots,b,\ldots,n)|^2 \to {\rm finite}.
\end{equation}
In other words, there is no singular contribution as $s_{ab} \to 0$,
and, when integrated over the small region of phase space where the
collinear approximation is valid, this colour ordered amplitudes
generates a negligible contribution to the cross section.
In eq.~(2.5),  $z$ is the fraction of the momentum carried by one of the gluons
and,
after integrating over the azimuthal angle of the plane containing the
collinear
particles with respect to the rest of the hard process,
the collinear splitting function
$P_{gg\to g}$ is given by,
\begin{equation}
P_{gg\to g} (z,s) = \frac{2}{s} P_{gg\to g} (z)
\label{eq:double}
\end{equation}
where the usual Altarelli-Parisi splitting kernel \cite{AP}
with the colour factor removed is given by,
\begin{equation}
P_{gg\to g} (z) = \left( \frac{1+z^4+(1-z)^4}{z(1-z)}\right).
\label{eq:pgg}
\end{equation}
In either soft or collinear limits, the process appears to
involve only $n-1$ hard gluons and the infrared singularities
must precisely cancel  \cite{BN,KLN}
with those from the $(n-1)$-gluon one-loop amplitudes.
Together, the real and virtual graphs form the next-to-leading order
perturbative contribution for infrared safe quantities associated
with the $(n-1)$ gluon scattering process.
Many methods have been developed to first isolate the infrared singularities
analytically and then numerically combine the remaining finite
real and virtual contributions \cite{fsks,slice,ert,subtract,hybrid}.

\section{Colour-ordered matrix elements squared}
\setcounter{equation}{0}
\label{sec:nnlo}

There are two five parton processes,
\begin{equation}
e^+e^- \to q\bar q ggg,
\label{eq:2q}
\end{equation}
and,
\begin{equation}
e^+e^- \to q\bar q Q\overline{Q} g,
\label{eq:4q}
\end{equation}
for which the lowest order matrix elements for process~(\ref{eq:2q})
can be written,
\begin{equation}
\M(Q_1;1,2,3;\overline{Q}_2) =
\widehat{\S}^5_\mu(Q_1;1,2,3;\overline{Q}_2)V^\mu,
\end{equation}
while for process~(\ref{eq:4q}) we have,
\begin{equation}
\M(Q_1,\overline{Q}_2;Q_3,\overline{Q}_4;1) =
\widehat{\T}^5_\mu(Q_1,\overline{Q}_2; Q_3,\overline{Q}_4;1)V^\mu.
\end{equation}
In these expressions, $V^\mu$ represents the lepton current, while
$\widehat{\S}^5_\mu$
and $\widehat{\T}^5_\mu$ are hadronic currents containing quarks and gluons.
The gluon colour is denoted by the adjoint representation label
$a_1,\ldots, a_3$ while that of the quark is $c_1,\ldots,c_4$.

The two quark-three gluon current $\widehat{\S}^5_\mu$
may be decomposed according to the colour
structure \cite{coldec},
\begin{equation}
\widehat{\S}^5_\mu(Q_1;1,2,3;\overline{Q}_2)
=ieg^3\sum_{P(1,2,3)} (T^{a_1}T^{a_2} T^{a_3})_{c_1c_2}
S_\mu(Q_1;1,2,3;\overline{Q}_2),
\label{eq:3gcur}
\end{equation}
where $\S_\mu(Q_1;1,2,3;\overline{Q}_2)$ represents the colourless
subamplitude where the gluons are emitted in an ordered way from the
quark line.
The colour matrices are normalised such that,
$$
{\rm Tr} \left(T^aT^b\right) = \frac{1}{2}\delta^{ab}.
$$
By summing over all permutations of gluon emission, all
Feynman diagrams and colour structures are accounted for.

The squared matrix elements for eq.~(\ref{eq:2q}) are simply,
\begin{eqnarray}
\lefteqn{\Big |\widehat{\S}^5_\mu V^\mu\Big |^2 =
e^2 \left(\frac{g^2N}{2}\right)^3
\left(\frac{N^2-1}{N}\right)}\nonumber \\
&\times&
\Bigg[ \sum_{P(1,2,3)} \left(
\Big |\S_\mu(Q_1;1,2,3;\overline{Q}_2) V^\mu\Big |^2 -
\frac{1}{N^2}
\Big |\S_\mu(Q_1;1,2,\tilde{3};\overline{Q}_2) V^\mu\Big |^2 \right ) \nonumber
\\
&&\hspace{1cm}+ \left(\frac{N^2+1}{N^4}\right)
\Big |\S_\mu(Q_1;\tilde{1},\tilde{2},\tilde{3};\overline{Q}_2) V^\mu\Big |^2
\Bigg ].
\end{eqnarray}
In the last two terms, the tilde indicates that that
gluon should be inserted in all positions in the amplitude.  In other words,
\begin{equation}
\S_\mu(Q_1;1,2,\tilde{3};\overline{Q}_2) =
\S_\mu(Q_1;1,2,3;\overline{Q}_2)+
\S_\mu(Q_1;1,3,2;\overline{Q}_2)+
\S_\mu(Q_1;3,1,2;\overline{Q}_2).
\end{equation}
In this case,  gluon 3 is effectively photon-like and the
contribution from the triple and quartic gluon vertices drops out.
For reference, the squared matrix elements for quark-antiquark production
together with a single gluon can be written,
\begin{equation}
 \Big |\widehat{\S}^3_\mu V^\mu\Big |^2 =
e^2 \left(\frac{g^2N}{2}\right)
\left(\frac{N^2-1}{N}\right)
\Big |\S_\mu(Q;G;\overline{Q}) V^\mu\Big |^2  .
\end{equation}

The four quark-one gluon current may be decomposed according to its
colour structure as follows,
\begin{eqnarray}
\lefteqn{
\widehat{\T}^5_\mu(Q_1,\overline{Q}_2;Q_3,\overline{Q}_4;1)
=i\frac{eg^3}{2}}\nonumber \\
&\times &  \Biggl [
T^{a_1}_{c_1c_4}\delta_{c_3c_2}
\T_\mu(Q_1,\overline{Q}_2;Q_3,\overline{Q}_4;1)
+ (Q_1 \leftrightarrow Q_3,\overline{Q}_2\leftrightarrow\overline{Q}_4)
- (Q_1 \leftrightarrow Q_3)
- (\overline{Q}_2\leftrightarrow\overline{Q}_4)
{}~\Biggr ],\nonumber \\
\end{eqnarray}
where the exchanges are understood to apply to the colour labels as well.
The colour ordered sub-current can be written,
\begin{eqnarray}
\T_\mu(Q_1,\overline{Q}_2;Q_3,\overline{Q}_4;1)
&=&
\delta_{Q_1Q_2}\delta_{Q_3Q_4}
\T^A_\mu(Q_1,\overline{Q}_2;Q_3,\overline{Q}_4;1)
\nonumber \\
&+&\frac{1}{N}
\delta_{Q_3Q_2}\delta_{Q_1Q_4}
\T^B_\mu(Q_1,\overline{Q}_2;Q_3,\overline{Q}_4;1),
\end{eqnarray}
where,
\begin{eqnarray}
\T^A_\mu(Q_1,\overline{Q}_2;Q_3,\overline{Q}_4;1)
&=&
\A_\mu^{Q_1Q_2}(Q_1;1;\overline{Q}_4|Q_3;\overline{Q}_2)
+\A_\mu^{Q_3Q_4}(Q_3;\overline{Q}_2|Q_1;1;\overline{Q}_4)
\nonumber \\
\T^B_\mu(Q_1,\overline{Q}_2;Q_3,\overline{Q}_4;1)
&=&  \B_\mu^{Q_1Q_4}(Q_1;1;\overline{Q}_4|Q_3;\overline{Q}_2)
+  \B_\mu^{Q_3Q_2}(Q_3;\overline{Q}_2|Q_1;1;\overline{Q}_4).
\end{eqnarray}
Here, $\delta_{Q_iQ_j}=1$ if quarks $i$ and $j$ have the same flavour.
The functions $\A^{Q_1Q_2}_\mu$ and $\B^{Q_1Q_2}_\mu$
describe Feynman diagrams where the
gauge boson couples to the ${Q_1Q_2}$ pair. However,
in $\A^{Q_1Q_2}_\mu$, the colour flows
along the gluon connecting the two quark pairs,
so that $Q_1$ and $\overline{Q}_4$ are the endpoints
of a colour antenna (and similarly $Q_3$ and $\overline{Q}_2$)
while in $\B^{Q_1Q_2}_\mu$, no colour is
transmitted between the quark pairs and
now $Q_1$ and $\overline{Q}_2$ form the endpoints
of a colour antenna (and similarly $Q_3$ and
$\overline{Q}_4$).
In each case, the gluon may be emitted from any position on the colour line.

Squaring the four quark-one gluon amplitude and summing over colours yields,
\begin{eqnarray}
\lefteqn{
\Big |\widehat{\T}^5_\mu V^\mu\Big |^2 = e^2 \left(
\frac{g^2N}{2}\right)^3\left(\frac{N^2-1}{N^2}\right)}\nonumber \\
&\times &
\Biggl [ ~
\Big |  \T_\mu(Q_1,\overline{Q}_2;Q_3,\overline{Q}_4;1)V^{\mu} \Big | ^2
+\Big |  \T_\mu(Q_1,\overline{Q}_4;Q_3,\overline{Q}_2;1)V^{\mu} \Big | ^2
\nonumber \\
&-& \frac{2}{N}\Re
\left(
\T_\mu(Q_1,\overline{Q}_2;Q_3,\overline{Q}_4;1)V^{\mu}
+\T_\mu(Q_3,\overline{Q}_4;Q_1,\overline{Q}_2;1)V^{\mu}
\right)
\left(\T_\mu(Q_1,\overline{Q}_4;Q_3,\overline{Q}_2;1)V^{\mu}\right)^{\dagger}
\Biggr ]\nonumber \\
&&+ (Q_1 \leftrightarrow Q_3,\overline{Q}_2\leftrightarrow\overline{Q}_4),
\end{eqnarray}
or equivalently,
\begin{eqnarray}
\lefteqn{
\Big |\widehat{\T}^5_\mu V^\mu\Big |^2 = e^2 \left(
\frac{g^2N}{2}\right)^3\left(\frac{N^2-1}{N^2}\right)}\nonumber \\
&\times &
\Biggl [ ~
\Big |  \T^A_\mu(Q_1,\overline{Q}_2;Q_3,\overline{Q}_4;1)V^{\mu} \Big | ^2
\nonumber \\
&&+\frac{1}{N^2}
\left(
\Big |  \T^B_\mu(Q_1,\overline{Q}_4;Q_3,\overline{Q}_2;1)V^{\mu} \Big | ^2
-
\Big |  \overline{\T}_\mu(Q_1,\overline{Q}_2;Q_3,\overline{Q}_4;1)V^{\mu} \Big
| ^2\right)
\nonumber \\
&&+ \frac{2\delta_{Q_1Q_3}}{N}\Re
\left (\T^A_\mu(Q_1,\overline{Q}_2;Q_3,\overline{Q}_4;1)V^{\mu} \right)
\left
(\T^B_\mu(Q_1,\overline{Q}_2;Q_3,\overline{Q}_4;1)V^{\mu}\right)^{\dagger}
\nonumber \\
&& - \frac{(N^2+1)}{2N^3}\delta_{Q_2Q_4}
\Re
\left(
\overline{\T}_\mu(Q_1,\overline{Q}_2;Q_3,\overline{Q}_4;1)V^{\mu}
\right)
\left(\overline{\T}_\mu(Q_1,\overline{Q}_4;Q_3,\overline{Q}_2;1)
V^{\mu}\right)^{\dagger}\nonumber \\
&&\hspace{1cm}+ (Q_1 \leftrightarrow
Q_3,\overline{Q}_2\leftrightarrow\overline{Q}_4)
+ \delta_{Q_1Q_3} (Q_1 \leftrightarrow Q_3)
+ \delta_{Q_2Q_4} (\overline{Q}_2\leftrightarrow\overline{Q}_4) \Biggr].
\end{eqnarray}
Here we have introduced the shorthand notation,
\begin{eqnarray}
\overline{\T}_\mu(Q_1,\overline{Q}_2;Q_3,\overline{Q}_4;1) &=&
\T^A_\mu(Q_1,\overline{Q}_2;Q_3,\overline{Q}_4;1) +
\T^A_\mu(Q_3,\overline{Q}_4;Q_1,\overline{Q}_2;1)\nonumber \\
&=&
\T^B_\mu(Q_1,\overline{Q}_4;Q_3,\overline{Q}_2;1) +
\T^B_\mu(Q_3,\overline{Q}_2;Q_1,\overline{Q}_4;1).
\end{eqnarray}
Note that in the case of identical quarks, there is an extra symmetry factor of
$1/4$ multiplying the matrix elements.

\newpage
\section{Colour unconnected double unresolved}
\setcounter{equation}{0}
\label{sec:unconnected}

In the cases where the two unresolved particles are not colour connected, the
factorisation of the amplitudes involves the well-known functions describing
single soft and collinear emission.
We first describe what is meant by colour connected and colour unconnected.

\subsection{Colour connection}

In the previous section we have seen how tree level matrix elements can be
decomposed
into colour ordered subamplitudes which have nice factorisation properties in
the infrared limits.   Therefore it is useful to view the matrix elements
in terms of the colour structure associated with the subamplitudes.   For
example,
in eq.~(\ref{eq:3gcur}), the two quark-three gluon subamplitude,
$S_\mu(Q_1;1,2,3;\overline{Q}_2)$ is associated with the colour structure
$(T^{a_1}T^{a_2} T^{a_3})_{c_1c_2}$.  This is a colour antenna that
ends on the quark/antiquark
colour charges $c_1$ and $c_2$ with ordered emission of gluons with
colour $a_1,\ldots,a_3$.
Within this colour antenna, gluon $1$ is colour connected to the quark
$Q_1$ and gluon 2, but not to the antiquark $\overline{Q}_2$ or to gluon 3.
In cases involving more than one quark-antiquark pair there can be many
colour antennae.
For example, the four quark amplitude
$\A_\mu^{Q_1Q_2}(Q_1;1;\overline{Q}_4|Q_3;\overline{Q}_2)$
of section 3 describes a process with two separate colour antennae.
In general, the particles in one antenna are not colour connected to
the particles
in one of the other antennae.   However, there is one case where particles
in adjoining antenna can usefully be thought of as colour connected.
This is when there is an antiquark at the end of one antenna and a like
flavour quark at the beginning of another,
$$
\A(\ldots,\overline{Q}|Q,\ldots).
$$
When this quark-antiquark pair are collinear, they combine to form a
gluon $G$, which then connects, or pinches together, the two separate
colour antennae, so that,
$$
|\A(\ldots,\overline{Q}|Q,\ldots )|^2 \to
P_{q\bar q \to G}(z,s_{Q\overline{Q}})|\A(\ldots,a,G,b,\ldots )|^2.
$$
A useful definition of colour ``connected'' therefore includes these
antennae pinching configurations along with the more straightforward
colour connection within a single antenna.
All other cases are colour ``unconnected''.

\subsection{Two collinear pairs}

Two pairs of particles may become collinear separately, but with the
particles in one or both
of the pairs themselves not colour ``connected''. 
In these cases, there are no
singular contributions containing both of the vanishing invariants.
For instance, if partons $\{a,d\}$ and $\{b,c\}$ are
collinear then,
\begin{equation}
|\A(\ldots,a,\ldots,b,\ldots,c,\ldots,d,\ldots)|^2 \to {\rm less~singular}.
\end{equation}
By this we mean there is no contribution proportional to $1/s_{ad}s_{bc}$ and
once again, when integrated over the small region of phase space relevant
for this approximation yields a negligible contribution.

The situation where
two pairs of colour ``connected'' particles are collinear is also rather
trivial.
If partons $a$ and $b$ form $P$, while $c$ and $d$ cluster to form $Q$,
so that $P$ and $Q$ are themselves colour unconnected, then,
\begin{equation}
|\A(\ldots,a,b,\ldots,c,d,\ldots)|^2 \to
 P_{ab\to P}(z_1,s_{ab})\,P_{cd\to Q}(z_2,s_{cd})\,
|\A(\ldots,P,\ldots,Q,\ldots)|^2.
\end{equation}
Here, $z_1$ and $z_2$ are the momentum fractions carried by $a$ and $c$
respectively.
A similar result holds if either of the pairs involves particles in
separate antennae, but which
are able to undergo antenna pinching.
The collinear splitting functions are related to the (colourless)
Altarelli-Parisi
splitting kernels by eq.~(\ref{eq:double})
which,
in the conventional dimensional regularisation
scheme \cite{CDR} with all particles
treated in  $d=4-2\epsilon$ dimensions,
are given by~\cite{AP},
\begin{eqnarray}
P_{qg \rightarrow q}(z)&=&
 \left( \frac{1+z^2-\epsilon(1-z)^2}{1-z} \right) , \\
P_{q\bar{q} \rightarrow g}(z)&=&
 \left( \frac{z^2+(1-z)^2-\epsilon}{1-\epsilon} \right),
\label{eq:pqg}
\end{eqnarray}
with $P_{gg \to g}$ given in eq.~(\ref{eq:pgg}) and
$P_{gq \to q}(z) = P_{qg \to q}(1-z)$.
As before, azimuthal averaging of the collinear particle plane is understood.

\subsubsection{Double collinear limit of $e^+e^- \to 5$~partons}

In this limit where two pairs of partons are simultaneously collinear,
the five parton matrix elements factorise
into the three parton matrix elements multiplied by a combination of
products of collinear splitting functions.
Summing over all possible unconnected double collinear limits, for the
two-quark currents we find,
\begin{eqnarray}
\Big |\S_\mu(Q_1;1,2,3;\overline{Q}_2) V^\mu\Big |^2 & \to &
\left(P_{Q_1 1 \to Q} P_{23 \to G} +
 P_{Q_1 1 \to Q} P_{3\overline{Q}_2 \to \overline{Q}}
+P_{12 \to G} P_{3\overline{Q}_2 \to \overline{Q}} \right )
\Big |\S^3_\mu V^\mu\Big |^2, \nonumber \\
\Big |\S_\mu(Q_1;1,2,\tilde{3};\overline{Q}_2) V^\mu\Big |^2 & \to &
\left(
P_{Q_1 1 \to Q} P_{2\overline{Q}_2 \to \overline{Q}}
+P_{Q_1 3 \to Q} P_{2\overline{Q}_2 \to \overline{Q}}
+P_{Q_1 1 \to Q} P_{3\overline{Q}_2 \to \overline{Q}} \right.\nonumber \\
&& \left .+ P_{Q_1 3 \to Q} P_{12 \to G}
+P_{12 \to G} P_{3\overline{Q}_2 \to \overline{Q}}\right )
\Big |\S^3_\mu V^\mu\Big |^2, \nonumber \\
\Big |\S_\mu(Q_1;\tilde{1},\tilde{2},\tilde{3};\overline{Q}_2) V^\mu\Big |^2
& \to &\sum_{P(1,2,3)}
P_{Q_1 1 \to Q} P_{2\overline{Q}_2 \to \overline{Q}}
\, \Big |\S^3_\mu V^\mu\Big |^2,
\end{eqnarray}
whilst the only contributing pieces for the four-quark process are,
\begin{eqnarray}
\Big | \T^A_\mu(Q_1,\overline{Q}_2;Q_3,\overline{Q}_4;1)V^{\mu} \Big |^2 & \to
&
\left( P_{Q_1 1 \to Q} P_{\overline{Q}_4 Q_3 \to G}
+ P_{1\overline{Q}_4\to \overline{Q}} P_{\overline{Q}_2 Q_1 \to G}
\right) \, \Big | \S^3_\mu V^\mu \Big |^2, \nonumber \\
\Big | \T^B_\mu(Q_1,\overline{Q}_4;Q_3,\overline{Q}_2;1)V^{\mu} \Big |^2 & \to
&
\left( P_{Q_1 1 \to Q} P_{\overline{Q}_4 Q_3 \to G}
+ P_{1\overline{Q}_2\to \overline{Q}} P_{\overline{Q}_4 Q_3 \to G}
\right) \, \Big | \S^3_\mu V^\mu \Big |^2, \nonumber \\
\Big | \overline{\T}_\mu(Q_1,\overline{Q}_2;Q_3,\overline{Q}_4;1)V^{\mu} \Big
|^2 & \to &
\left( P_{Q_1 1 \to Q} P_{\overline{Q}_4 Q_3 \to G}
+ P_{1\overline{Q}_2\to \overline{Q}} P_{\overline{Q}_4 Q_3 \to G}
\right. \nonumber \\ && \left.
+P_{Q_3 1 \to Q} P_{\overline{Q}_2 Q_1 \to G}
+ P_{1\overline{Q}_4\to \overline{Q}} P_{\overline{Q}_2 Q_1 \to G}
\right) \, \Big | \S^3_\mu V^\mu \Big |^2.
\end{eqnarray}
For brevity we have dropped the arguments of the splitting functions.
Explicitly, for $n_F$ flavours of quark, we find,
\begin{eqnarray}
\lefteqn{\frac{1}{3!} \Big |\widehat{\S}^5_\mu V^\mu\Big |^2
+
\Big |\widehat{\T}^5_\mu V^\mu\Big |^2  = \left(\frac{g^2N}{2}\right)^2
\left(\frac{N^2-1}{N^2}\right)
\Big |\widehat{\S}^3_\mu V^\mu\Big |^2}\nonumber \\
&\times &\Biggl(
 P_{Q_1 1\to Q}P_{23 \to G} +
 P_{3\overline{Q}_2\to \overline{Q}}P_{12 \to G}
+P_{Q_1 1\to Q}P_{3\overline{Q}_2\to \overline{Q}}
\nonumber \\
&&
+ \frac{2n_F}{N}
P_{Q_1 1 \to Q} P_{\overline{Q}_4 Q_3 \to G}
- \frac{1}{2N^2}
\left(
P_{Q_1 1\to Q} P_{2\overline{Q}_2\to \overline{Q}}
+ P_{Q_1 3\to Q} P_{2\overline{Q}_2\to \overline{Q}}
\right)
\Biggr).
\end{eqnarray}
Note also that the identical gluon factor $1/3!$ is eliminated since
each term in the sum over permutations produces an identical contribution.
It is interesting to review the origin of the factor of $n_F$.  There are
$n_F(n_F-1)/2$ contributions from two unlike pairs of quarks, each of which
generates two sets of singular limits - that indicated plus the
symmetric term $(Q_1 \leftrightarrow
Q_3,\overline{Q}_2\leftrightarrow\overline{Q}_4)$.
In addition there are $n_F$ like-quark pair contributions, which after the
symmetries have been applied yield four singular limits.  However, the
identical quark contribution is multiplied by the identical particle
factor $1/4$ so that the net result is,
$$
\frac{n_F(n_F-1)}{2} \times 2 + n_F \times 4 \times \frac{1}{4} = n_F^2.
$$
One factor of $n_F$ is absorbed into the three parton matrix elements
$\Big |\widehat{\S}^3_\mu V^\mu\Big |^2$, while the other appears
as an explicit factor.

\subsection{Triple collinear factorisation}

If three collinear particles are  colour ``unconnected'' then there is
no singularity. So if $a$, $b$ and $c$ all become collinear,
\begin{equation}
|\A(\ldots,a,\ldots,b,\ldots,c,\ldots)|^2 \to {\rm finite},
\end{equation}
and there is no singular contribution involving the invariants
$s_{ab}$, $s_{bc}$ or $s_{abc}$. As before, because the region of phase
space where the triple collinear limit is valid is extremely small,
this gives a negligible contribution to the cross section.
When two of the three collinear particles are colour ``connected'' we find a
singular result,
\begin{equation}
|\A(\ldots,a,\ldots,b,c,\ldots)|^2 \to 1/s_{bc}.
\end{equation}
However, when integrated over the triple collinear region of phase space that
requires
$s_{ab}$, $s_{bc}$ or $s_{abc}$ all to be small,  we again obtain a
negligible contribution
that is proportional to the small parameter defining the extent of
the triple collinear phase space.
We therefore ignore contributions of this type.

\subsection{Soft/collinear factorisation}

Two particles may be unresolved if one of them is a soft gluon and another
pair are collinear. When the soft gluon $g$ is not colour connected to either
of the
colour ``connected'' collinear particles $c$ and $d$,
factorisation is straightforward,
\begin{equation}
|\A(\ldots,a,g,b,\ldots,c,d,\ldots)|^2 \to
S_{agb}(s_{ab},s_{ag},s_{bg})P_{cd\to P}(z,s_{cd})\,
|\A(\ldots,a,b,\ldots,P,\ldots)|^2.
\end{equation}

\subsubsection{Soft/collinear limit of $e^+e^- \to 5$~partons}

In the soft/collinear limit, the five parton matrix elements again
factorise
into a singular factor multiplying the squared two-quark current relevant
for three parton production,
\begin{eqnarray}
\Big |\S_\mu(Q_1;1,2,3;\overline{Q}_2) V^\mu\Big |^2 & \to &
\left(S_{Q_1 1 2}P_{3\overline{Q}_2 \to \overline{Q}}
+ P_{Q_1 1 \to Q}S_{23 \overline{Q}_2} \right)
 \Big |\S^3_\mu V^\mu\Big |^2, \nonumber \\
\Big |\S_\mu(Q_1;1,2,\tilde{3};\overline{Q}_2) V^\mu\Big |^2 & \to &
\left(S_{Q_1 1 2}P_{3\overline{Q}_2 \to \overline{Q}}
+ P_{Q_1 3 \to Q}S_{12 \overline{Q}_2}
+ P_{12 \to G}  S_{Q_1 3 \overline{Q}_2}
\right)
 \Big |\S^3_\mu V^\mu\Big |^2. \nonumber \\
\end{eqnarray}
Note that for $\Big |\S_\mu(Q_1;\tilde{1},\tilde{2},\tilde{3};\overline{Q}_2)
V^\mu\Big |^2 $, the
soft and collinear limits are considered to be overlapping and will be dealt
with in section~\ref{subsec:softcol}.

In the four-quark current case, the soft/collinear limit has only two
colour-unconnected contributions. The first is given by,
\begin{equation}
\Big | \T^B_\mu(Q_1,\overline{Q}_4;Q_3,\overline{Q}_2;1)V^{\mu} \Big |^2 \to
S_{Q_1 1 \overline{Q}_2}P_{Q_3\overline{Q}_4 \to G}
\, \Big | \S^3_\mu V^\mu \Big |^2,
\end{equation}
whilst the limit of,
$\Big | \overline{\T}_\mu(Q_1,\overline{Q}_2;Q_3,\overline{Q}_4;1)
V^{\mu} \Big |^2,$
again involves both unconnected and connected factors and therefore discussion
of this will also be deferred until section~\ref{subsec:softcol}. The other
subamplitudes
vanish in the unconnected soft/collinear limit.

Applying these limits to the full five parton matrix elements is straightfoward
and, after removing identical particle factors where necessary,
we find,
\begin{eqnarray}
\lefteqn{\frac{1}{3!} \Big |\widehat{\S}^5_\mu V^\mu\Big |^2
+
\Big |\widehat{\T}^5_\mu V^\mu\Big |^2  = \left(\frac{g^2N}{2}\right)^2
\Big |\widehat{\S}^3_\mu V^\mu\Big |^2}\nonumber \\
&\times &\Biggl[
\left(\frac{N^2-1}{N^2}\right)
\left(S_{Q_1 1 2}P_{3\overline{Q}_2 \to \overline{Q}}
+ P_{Q_1 1 \to Q}S_{23 \overline{Q}_2} \right)
- \frac{1}{N^2}
 P_{12 \to G}  S_{Q_1 3 \overline{Q}_2}
\nonumber \\
&&
+ \frac{n_F}{N^3}
 S_{Q_1 1 \overline{Q}_2}P_{Q_3\overline{Q}_4 \to G}
\Biggr].
\end{eqnarray}

\subsection{Two soft gluons}

When two unconnected gluons are soft, the factorisation is again simple
\cite{multsoft}.
For gluons $g_1$ and $g_2$ soft we find,
\begin{eqnarray}
|\A(\ldots,a,g_1,b,\ldots,c,g_2,d,\ldots)|^2 & \to &
 S_{ag_1b}(s_{ab},s_{ag_1},s_{bg_1})S_{cg_2d}(s_{cd},s_{cg_2},s_{dg_2})
\nonumber \\ && \times |\A(\ldots,a,b,\ldots,c,d,\ldots)|^2,
\end{eqnarray}
so that the singular factor is merely the product of two single soft gluon
emission factors given by eq.~(\ref{eq:ssoft}). Note that $b=c$ is allowed.

\subsubsection{Double soft limit of $e^+e^- \to 5$~partons}

The sum over the unconnected double soft limits of the colour ordered
subamplitudes
can be easily read off,
\begin{eqnarray}
\Big |\S_\mu(Q_1;1,2,3;\overline{Q}_2) V^\mu\Big |^2 & \to &
S_{Q_1 1 2}S_{23\overline{Q}_2}
 \Big |\S^3_\mu V^\mu\Big |^2, \nonumber \\
\Big |\S_\mu(Q_1;1,2,\tilde{3};\overline{Q}_2) V^\mu\Big |^2 & \to &
\left(S_{Q_1 1 2}S_{Q_1 3 \overline{Q}_2}
+ S_{Q_1 3 \overline{Q}_2}S_{12 \overline{Q}_2}
\right)
 \Big |\S^3_\mu V^\mu\Big |^2, \nonumber \\
\Big |\S_\mu(Q_1;\tilde{1},\tilde{2},\tilde{3};\overline{Q}_2) V^\mu\Big |^2 &
\to &
 \frac{1}{2} \sum_{P(1,2,3)} S_{Q_1 1 \overline{Q}_2}S_{Q_1 2 \overline{Q}_2}
\Big |\S^3_\mu V^\mu\Big |^2.
\end{eqnarray}
There is no contribution from the four-quark matrix elements.
Inserting these limits into the full five parton matrix elements yields,
\begin{eqnarray}
\lefteqn{\frac{1}{3!} \Big |\widehat{\S}^5_\mu V^\mu\Big |^2
+
\Big |\widehat{\T}^5_\mu V^\mu\Big |^2  =
\left(\frac{g^2N}{2}\right)^2
\Big |\widehat{\S}^3_\mu V^\mu\Big |^2}\nonumber \\
&\times&
\Bigg[ S_{Q_1 1 2}S_{23\overline{Q}_2} -
\frac{1}{N^2}
\left(S_{Q_1 1 2}S_{Q_1 3 \overline{Q}_2}
+ S_{Q_1 3 \overline{Q}_2}S_{12 \overline{Q}_2}
\right)
\nonumber \\
&&\hspace{1cm}+ \left(\frac{N^2+1}{2N^4}\right)
 S_{Q_1 1 \overline{Q}_2}S_{Q_1 2 \overline{Q}_2}
\Bigg ],
\end{eqnarray}
where once again the sum over permutations is eliminated by the identical
particle factor.

\newpage
\section{Colour connected double unresolved}
\setcounter{equation}{0}
\label{sec:connected}

The factorisation that occurs when the two unresolved particles are colour
``connected''
is necessarily more involved than that in section~\ref{sec:unconnected}.
In particular, we will need to intoduce new functions to describe this
factorisation.

\subsection{Triple collinear factorisation}
\label{subsec:triple}

When three colour ``connected'' particles cluster to form a single parent
parton there are four basic clusterings,
\begin{eqnarray*}
&& ggg \rightarrow G, \qquad qgg \rightarrow Q, \\
&& g{\bar q}q \rightarrow G, \qquad q{\bar q}q \rightarrow Q,
\end{eqnarray*}
and the colour ordered sub-amplitude
squared for an $n$-parton process then factorises in
the triple collinear limit,
\begin{equation}
|\A(\ldots,a,b,c,\ldots)|^2 \rightarrow
P_{abc \rightarrow P}
|\A(\ldots,P,\ldots)|^2.
\end{equation}
As before, partons able to undergo antenna pinching are considered to be
colour connected,
so that there may be contributions from amplitudes such as
$\A(\ldots,a,b|c,\ldots)$.
The triple collinear splitting function for partons $a$, $b$ and
$c$ clustering to form the parent parton $P$ is generically,
\begin{equation}
P_{abc \rightarrow P}(w,x,y,s_{ab},s_{ac},s_{bc},s_{abc}),
\end{equation}
where $w$, $x$ and $y$ are the momentum fractions of the clustered partons,
\begin{equation}
\label{momfrac}
p_a=wp_P, \qquad p_b=xp_P, \qquad p_c=yp_P, \qquad \mbox{with } w+x+y=1.
\end{equation}
In addition to depending on the momentum fractions carried by the
clustering partons, the splitting function also depends on the
invariant masses
of parton-parton pairs and the invariant mass of the whole cluster.
In this respect, they are different from the splitting functions
derived in the jet-calculus approach \cite{jetcalculus}, and
implemented in the
shower Monte Carlo NLLJET \cite{NLLJET}, which depend only on
the momentum fractions.

The triple collinear splitting functions  $P_{abc \rightarrow P}$
are obtained by retaining terms in the full matrix
element squared that possess two of the `small' denominators $s_{ab}$,
$s_{ac}$,
$s_{bc}$ and $s_{abc}$.
As before, we consider the explicit forms of the
$\gamma^* \to $ four and five parton squared matrix elements
and work in conventional dimensional regularisation, with all external
particles in $d=4-2\e$ dimensions.
Similar results could be derived using helicity methods or by examining the
on-shell limits of the recursive gluonic and quark currents of
ref.~\cite{current}.

Although the splitting functions are universal, and apply to any
process involving the same three colour connected particles,
for processes involving spin-1 particles, there are additional
(non-universal) azimuthal correlations due to rotations of the
polarisation vectors.
These angular correlations do not contribute to the underlying
infrared singularity structure and vanish after all azimuthal integrations
have been carried out and we therefore systematically omit them.

A further check
on our results is provided by the strong-ordered limit, where the particles
become collinear sequentially rather than at the same time. In this limit
these functions factorise into the product of two usual
collinear splitting functions plus azimuthal terms,
agreeing with the results of~\cite{Knowles}.

\subsubsection{Three collinear gluons}

Firstly, examining the sub-amplitudes for multiple gluon scattering,
we find
that the colour-ordered function $P_{ggg \rightarrow G}$ is given by,
\begin{eqnarray}
\lefteqn{P_{abc \rightarrow G}(w,x,y,
 s_{ab},s_{bc},s_{abc}) =
 8 \times \Biggl\{ } \nonumber \\
 &+& \frac{(1-\e)}{s_{ab}^2s_{abc}^2}\frac{(xs_{abc}-(1-y)s_{bc})^2}{(1-y)^2}
    +\frac{2(1-\e)s_{bc}}{s_{ab}s_{abc}^2} +\frac{3(1-\e)}{2s_{abc}^2}\nonumber
\\
 &+& \frac{1}{s_{ab}s_{abc}} \left(
     \frac{(1-y(1-y))^2}{yw(1-w)}-2\frac{x^2+xy+y^2}{1-y}
    +\frac{xw-x^2y-2}{y(1-y)} +2\e \frac{x}{(1-y)} \right)\nonumber \\
 &+& \frac{1}{2s_{ab}s_{bc}} \left(
    3x^2 - \frac{2(2-w+w^2)(x^2+w(1-w))}{y(1-y)}
    + \frac{1}{yw} + \frac{1}{(1-y)(1-w)} \right) \Biggr\} \nonumber \\
 &+& ( s_{ab} \leftrightarrow s_{bc}, w \leftrightarrow y)
    + {\rm azimuthal~terms}.
\end{eqnarray}
This splitting function is symmetric under the exchange of the outer gluons,
and contains poles only in $s_{ab}$ and $s_{bc}$.

\subsubsection{Two gluons with a collinear quark or antiquark}

There are  two distinct splitting functions representing the clustering of two
gluons
and a quark which depend on whether or not the gluons are symmetrised over.
In the unsymmetrised case, there will be poles in $s_{g_1g_2}$, due
to contributions from the triple gluon vertex which are not present
in the QED-like case.
For the pure QCD splitting we find,
\begin{eqnarray}
\lefteqn{P_{qg_1g_2 \rightarrow Q}
 (w,x,y,s_{qg_1},s_{qg_2},s_{g_1g_2},s_{qg_1g_2}) =
 4 \times \Biggl\{ } \nonumber \\
 &+&\frac{1}{s_{qg_1}s_{g_1g_2}} \left( (1-\e) \left(
  \frac{1+w^2}{y}+\frac{1+(1-y)^2}{(1-w)} \right)
  +2\e \left( \frac{w}{y}+\frac{1-y}{1-w} \right) \right) \nonumber \\
 &+&\frac{1}{s_{qg_1}s_{qg_1g_2}} \left( (1-\e) \left( \frac{
  (1-y)^3+w(1-x)-2y}{y(1-w)} \right)
  - \e \left( \frac{2(1-y)(y-w)}{y(1-w)} -x \right) -\e^2 x \right) \nonumber
\\
 &+&\frac{1}{s_{g_1g_2}s_{qg_1g_2}} \left( (1-\e) \left( \frac{
  (1-y)^2 (2-y)+x^3+2xw-2-y}{y(1-w)} \right) +2\e \frac{(xw-y-2yw)}{y(1-w)}
\right) \nonumber \\
 &+&(1-\e) \left( \frac{2\left( x{s_{qg_1g_2}}-(1-w)s_{qg_1} \right)^2}
  {s_{g_1g_2}^2s_{qg_1g_2}^2(1-w)^2}
  +\frac{1}{s_{qg_1g_2}^2} \left( 4\frac{s_{qg_1}}{s_{g_1g_2}}
   +(1-\e) \frac{s_{g_1g_2}}{s_{qg_1}} + (3-\e) \right) \right) \Biggr\},
\end{eqnarray}
while for  the QED-like splitting where one or other or both gluons in the
colour ordered amplitude are symmetrised over,
\begin{eqnarray}
\lefteqn{P_{q\tilde{g_1}\tilde{g_2} \rightarrow Q}
 (w,x,y,s_{qg_1},s_{qg_2},s_{qg_1g_2}) =
 4 \times \Biggl\{ } \nonumber \\
 &+&\frac{1}{2s_{qg_1}s_{qg_2}} \frac{w}{xy}
\left( 1+w^2-\e(x^2+xy+y^2)-\e^2 xy \right) \nonumber \\
 &+&\frac{1}{s_{qg_1}s_{qg_1g_2}} \frac{1}{xy} \left( w(1-x+\e^2
xy)+(1-y)^3-\e(1-y)
  (x^2+xy+y^2)+\e^2 xy \right) \nonumber \\
 &-&\frac{(1-\e)}{s_{qg_1g_2}^2} \left( (1-\e)
  \frac{s_{qg_1}}{s_{qg_2}} - \e \right) \Biggr\}
  + ( s_{qg_1} \leftrightarrow s_{qg_2}, x \leftrightarrow y) .
\end{eqnarray}
The function $P_{q\tilde{g_1}\tilde{g_2} \rightarrow Q}$ can be interpreted
as the relevant triple collinear splitting function with one or both of the
gluons replaced by photons. As such, this result echoes that
found in~\cite{Aude} for $P_{qg\gamma \rightarrow Q}$.
Using charge conjugation, we see that the functions representing clustering
of two gluons with an antiquark are simply,
\begin{eqnarray}
P_{g_1 g_2 \bar{q} \rightarrow \overline{Q}}
 (w,x,y,s_{g_1g_2},s_{g_2\bar{q}},s_{g_1 \bar{q}},s_{g_1 g_2\bar{q}}) &=&
P_{q g_1g_2 \rightarrow Q}
 (y,x,w,s_{\bar{q}g_2},s_{\bar{q}g_1},s_{g_1 g_2},s_{g_1
g_2\bar{q}}),\nonumber\\
P_{\tilde{g_1}\tilde{g_2} \bar{q} \rightarrow \overline{Q}}
 (w,x,y,s_{g_2\bar{q}},s_{g_1 \bar{q}},s_{g_1 g_2\bar{q}}) &=&
P_{q \tilde{g_1}\tilde{g_2} \rightarrow Q}
 (y,x,w,s_{\bar{q}g_2},s_{\bar{q}g_1},s_{g_1 g_2\bar{q}}).
\end{eqnarray}

\subsubsection{A quark-antiquark pair with a collinear gluon}

Similarly the clustering of a gluon with a quark-antiquark pair into
a parent gluon again has two distinct functions. For example, there is a
singular contribution from the four quark matrix elements when
$\overline{Q}_4$, $Q_3$ and the gluon
cluster,
\begin{eqnarray*}
\Big | \T^A_\mu(Q_1,\overline{Q}_2;Q_3,\overline{Q}_4;1)V^{\mu} \Big |^2
&=&
\Big |  \A_\mu^{Q_1Q_2}(Q_1;1;\overline{Q}_4|Q_3;\overline{Q}_2)V^{\mu}
+\A_\mu^{Q_3Q_4}(Q_3;\overline{Q}_2|Q_1;1;\overline{Q}_4)V^{\mu}\Big |^2
\\&\to&
P_{1\overline{Q}_4Q_3 \to G} \, \Big | \S^3_\mu V^\mu \Big |^2.
\end{eqnarray*}
Here, the singular contribution comes entirely from the first $\A$ term
where
combining $\overline{Q}_4$, $Q_3$  pinches the two colour lines together.
Explicit evaluation yields,
\begin{eqnarray}
\lefteqn{P_{g{\bar q}q \rightarrow G}
 (w,x,y,s_{g\bar q},s_{\bar q q},s_{g{\bar q}q}) =
 4 \times \Biggl\{ } \nonumber \\
&-& \frac{1}{s_{g{\bar q}q}^2} \left( 4\frac{s_{g{\bar q}}}{s_{\bar q q}}
 +(1-\e)\frac{s_{\bar q q}}{s_{g{\bar q}}} + (3-\e) \right)
 - \frac{2 \left( xs_{g{\bar q}q}-(1-w)s_{g{\bar q}} \right)^2}{s_{\bar q q}^2
  s_{g{\bar q}q}^2(1-w)^2} \nonumber \\
&+& \frac{1}{s_{g{\bar q}}s_{g{\bar q}q}} \left( \frac{(1-y)}{w(1-w)}-y-2w-\e
 -\frac{2x(1-y)(y-w)}{(1-\e)w(1-w)} \right) \nonumber \\
&+&  \frac{1}{s_{g{\bar q}}s_{\bar q q}} \left( \frac{x\left( (1-w)^3-w^3
\right)}
 {w(1-w)} -\frac{2x^2 \left( 1-yw-(1-y)(1-w) \right)}{(1-\e)w(1-w)} \right)
\nonumber \\
&+& \frac{1}{s_{{\bar q}q}s_{g{\bar q}q}} \left( \frac{(1+w^3+4xw)}{w(1-w)}
 +\frac{2x \left( w(x-y)-y(1+w) \right)}{(1-\e)w(1-w)} \right) \Biggr\}
+ {\rm azimuthal~terms}.
\end{eqnarray}
Again applying charge conjugation yields the further relation,
\begin{equation}
P_{{\bar q}qg \rightarrow G}
 (w,x,y,s_{\bar{q}q},s_{qg},s_{{\bar q}qg}) =
P_{g{\bar q}q \rightarrow G}
 (y,x,w,s_{gq},s_{\bar q q},s_{{\bar q}qg}),
\end{equation}
describing instances where the gluon is colour connected to the quark
rather than the antiquark.

There is a further contribution when the quark-antiquark and gluon
combine to form a photon-like colour singlet.
This occurs when,
\begin{eqnarray*}
\Big | \T^B_\mu(Q_3,\overline{Q}_2;Q_1,\overline{Q}_4;1)V^{\mu} \Big |^2
&=&
\Big | \B_\mu^{Q_3Q_4}(Q_3;1;\overline{Q}_4|Q_1;\overline{Q}_2)V^{\mu}
+  \B_\mu^{Q_1Q_2}(Q_1;\overline{Q}_2|Q_3;1;\overline{Q}_4)V^{\mu} \Big |^2
\\
&\to&
P_{Q_3 \tilde{1} \overline{Q}_4  \to \tilde{G}}
\, \Big | \S^3_\mu V^\mu \Big |^2.
\end{eqnarray*}
In this case the singular contribution is produced by the second $\B$ term and
is due to the entire $Q_3;1;\overline{Q}_4$ antenna pinching to form a gluon
which is then inserted in a symmetrised way (i.e. with a tilde)
into the remaining colour antenna.
This QED-like splitting function is given by,
\begin{eqnarray}
\lefteqn{P_{q g {\bar q} \rightarrow \tilde{G}}
 (w,x,y,s_{qg},s_{g\bar q},s_{\bar q q},s_{qg{\bar q}}) =
 4 \times \Biggl\{ } \nonumber \\
&-& \frac{1}{s_{qg{\bar q}}^2} \left( (1-\e)
 \frac{s_{q{\bar q}}}{s_{qg}} +1 \right)
 + \frac{1}{s_{g\bar q}s_{qg}} \left( (1+x^2)-\frac{x+2wy}{1-\e} \right)
\nonumber \\
&-& \frac{1}{s_{qg}s_{qg{\bar q}}} \left( 1+2x+\e-\frac{2(1-y)}{(1-\e)} \right)
  \Biggr\} + ( s_{qg} \leftrightarrow s_{g\bar q}, w \leftrightarrow y)
+ {\rm azimuthal~terms}.
\end{eqnarray}

\subsubsection{A quark-antiquark pair with a collinear quark or antiquark}

Lastly, we consider the clustering of a quark-antiquark
pair~($Q{\overline{Q}}$)
and a quark~($q$) to form a parent quark $q^\prime$ with the same flavour as
$q$.
The splitting function depends upon
whether or not the quarks are identical,
\begin{eqnarray}
\label{qQQQ}
P_{q{\overline{Q}}Q \rightarrow q^\prime} =
 P_{q{\overline{Q}}Q \rightarrow q^\prime}^{{\rm non-ident.}}
-\frac{\delta_{qQ}}{N} P_{q{\overline{Q}}Q \rightarrow q^\prime}^{{\rm
ident.}},
\end{eqnarray}
where $\delta_{qQ}=1$ for identical quarks.
If quarks $Q_1$, $Q_3$ and $\overline{Q}_4$ are clustered to form $Q$,
then we find a non-identical quark contribution,
\begin{eqnarray*}
\Big | \T^A_\mu(Q_3,\overline{Q}_4;Q_1,\overline{Q}_2;1)V^{\mu} \Big |^2
&=&
\Big |  \A_\mu^{Q_3Q_4}(Q_3;1;\overline{Q}_2|Q_1;\overline{Q}_4)V^{\mu}
+\A_\mu^{Q_1Q_2}(Q_1;\overline{Q}_4|Q_3;1;\overline{Q}_2)V^{\mu}\Big |^2
\\&\to&
P_{Q_1\overline{Q}_4Q_3 \to Q}^{{\rm non-ident.}}
\, \Big | \S^3_\mu V^\mu \Big |^2,
\end{eqnarray*}
with,
\begin{eqnarray}
\lefteqn{P_{q\overline{Q}Q \rightarrow q^\prime}^{{\rm non-ident.}}(w,x,y,
 s_{q\overline{Q}},s_{Q\overline{Q}},s_{qQ\overline{Q}}) =
 4 \times \Biggl\{ - \frac{1}{s_{qQ\overline{Q}}^2} \left( (1-\e)
  +\frac{2s_{q\overline{Q}}}{s_{Q\overline{Q}}} \right) } \nonumber \\
 && -\frac{2\left( x s_{qQ\overline{Q}}-(1-w) s_{q\overline{Q}} \right)^2}
  {s_{Q\overline{Q}}^2s_{qQ\overline{Q}}^2(1-w)^2}
  +\frac{1}{s_{Q\overline{Q}}s_{qQ\overline{Q}}} \left(
  \frac{1+x^2+(x+w)^2}{(1-w)} - \e(1-w) \right) \Biggr\}.
\end{eqnarray}
The singular contribution is now generated by the square of the second $\A$
term;
pinching $Q_3$ and $\overline{Q}_4$ together connects the two colour antennae
together and combining with $Q_1$ ensures that the vector boson couples to
a flavour singlet $Q_1 \overline{Q}_2$ pair.
Precisely the same function describes the triple collinear limit of
the $\T^B$ functions.  We find that in the same limit,
\begin{eqnarray*}
\Big | \T^B_\mu(Q_1,\overline{Q}_4;Q_3,\overline{Q}_2;1)V^{\mu} \Big |^2
&=&
\Big |
\B_\mu^{Q_1Q_2}(Q_1;1;\overline{Q}_2|Q_3;\overline{Q}_4) V^{\mu}
+  \B_\mu^{Q_3Q_4}(Q_3;\overline{Q}_4|Q_1;1;\overline{Q}_2) V^{\mu}\Big |^2
\\&\to&
P_{Q_1\overline{Q}_4Q_3 \to Q}^{{\rm non-ident.}}
\, \Big | \S^3_\mu V^\mu \Big |^2,
\end{eqnarray*}
where this time the first term alone contributes. Here $Q_3$ and
$\overline{Q}_4$ combine to form a photon which clusters with $Q_1$.
As before, charge conjugation generates the associated function for
an antiquark combining with a quark-antiquark pair,
\begin{equation}
P_{\overline{q}Q\overline{Q} \rightarrow \overline{q}^\prime}^{{\rm
non-ident.}}(w,x,y,
 s_{\overline{q}Q},s_{Q\overline{Q}},s_{\overline{q}Q\overline{Q}})
=
P_{q\overline{Q}Q \rightarrow q^\prime}^{{\rm non-ident.}}(w,x,y,
 s_{q\overline{Q}},s_{Q\overline{Q}},s_{qQ\overline{Q}}).
\end{equation}

When the flavours of the clustering quarks are the same, there is an
additional contribution coming from the interference
terms of the four-quark matrix elements. For instance, when $Q_1$,
$\overline{Q}_4$ and $Q_3$ combine,
\begin{eqnarray*}
\lefteqn{\Re
\left(
\T^A_\mu(Q_3,\overline{Q}_4;Q_1,\overline{Q}_2;1)V^{\mu}
\right)
\left(\T^B_\mu(Q_3,\overline{Q}_4;Q_1,\overline{Q}_2;1)V^{\mu}\right)^{\dagger}
}\\
&\sim&
\A_\mu^{Q_3Q_4}(Q_3;1;\overline{Q}_2|Q_1;\overline{Q}_4)V^{\mu}
\left(\B_\mu^{Q_3Q_2}(Q_3;1;\overline{Q}_2|Q_1;\overline{Q}_4)V^{\mu}
\right)^\dagger
\\
&\to& - \frac{1}{2} P_{Q_1\overline{Q}_4 Q_3 \to Q}^{\rm ident.}
\, \Big | \S^3_\mu V^\mu \Big |^2,
\end{eqnarray*}
where,
\begin{eqnarray}
\lefteqn{P_{q\overline{Q}Q \rightarrow q^\prime}^{{\rm ident.}}(w,x,y,
 s_{q\overline{Q}},s_{Q\overline{Q}},s_{qQ\overline{Q}}) =  4 \times \Biggl\{ }
\nonumber \\
 &-& \frac{(1-\e)}{s_{qQ\overline{Q}}^2} \left(
  \frac{2s_{q\overline{Q}}}{s_{Q\overline{Q}}}+2+\e \right)
  -\frac{1}{2s_{q\overline{Q}}s_{Q\overline{Q}}} \left(
  \frac{x(1+x^2)}{(1-y)(1-w)}-\e x \left(
  \frac{2(1-y)}{(1-w)}+1+\e \right) \right)
\nonumber \\
 &+& \frac{1}{s_{Q\overline{Q}}s_{qQ\overline{Q}}} \left(
  \frac{1+x^2}{(1-y)}+\frac{2x}{(1-w)} - \e \left(
  \frac{(1-w)^2}{(1-y)}+(1+x)+\frac{2x}{(1-w)}+\e(1-w) \right) \right) \Biggr\}
\nonumber \\
 &+& ( s_{q\overline{Q}} \leftrightarrow s_{Q\overline{Q}}, y \leftrightarrow
w).
\end{eqnarray}
Here, there are poles in the matrix elements when $\overline{Q}_4$
clusters with both $Q_3$ and $Q_1$ and the triple collinear function is
symmetric under $q \leftrightarrow Q$.

\subsubsection{The $N=1$ SUSY identity}

These triple-collinear splitting functions, like the ordinary
Altarelli-Parisi splitting kernels, can be related by means of an $N=1$
supersymmetry identity.
In unbroken supersymmetric theories, the masses of gluon and gluino are
identical thereby ensuring that
the self-energies of the two particles are equal.  Considering
all two particle cuts of the one-loop diagrams contributing
to these self-energies, then in terms of the colour stripped Altarelli-Parisi
kernels
eq.~\ref{eq:pgg} and eq.~\ref{eq:pqg},
\begin{equation}
P_{gg \rightarrow g}(z)+P_{q \bar q \rightarrow g}(z)
 = P_{qg \rightarrow q}(z)+P_{gq \rightarrow q}(z).
\end{equation}
Here the quark plays
the role of the gluino.
Note that in conventional
dimensional regularisation the number of degrees of
freedom for the gluon and gluino are no longer equal and the supersymmetry
is broken\footnote{We note that there are other variants of dimensional
regularisation where the gluon and gluino degrees of freedom are equal.}.
Therefore this result is not true away from four
dimensions.
Similarly, by considering the three particle cuts of the
relevant two-loop contributions to the self energies,
and omitting the invariants in the arguments of the functions, we find,
\begin{eqnarray}
\lefteqn{
 \sum_{P(a,b,c)} \Biggl( P_{ggg \rightarrow G}(a,b,c)
+ 2 P_{g\bar q q \rightarrow G}(a,b,c)
+ P_{q \tilde{g}\bar q  \rightarrow G}(a,b,c) \Biggr) = } \\
&& \sum_{P(a,b,c)} \Biggl( 2 P_{qgg \rightarrow Q}(a,b,c)
+ P_{q\tilde{g}\tilde{g} \rightarrow Q}(a,b,c)
+ 2 P_{q\bar q q \rightarrow Q}^{\rm non-ident.}(a,b,c)
+ P_{q\bar q q \rightarrow Q}^{\rm ident.}(a,b,c) \Biggr), \nonumber
\end{eqnarray}
provided we set $\e = 0$. This non-trivial relation between the
splitting functions is a further check on the results presented in this
section.

\subsubsection{Triple collinear limit of $e^+e^- \to 5$~partons}

Examining the two-quark currents in the triple collinear limit is now
straightforward.
Summing over all triple collinear limits, we find,
\begin{eqnarray}
\Big |\S_\mu(Q_1;1,2,3;\overline{Q}_2) V^\mu\Big |^2 & \to &
\left( P_{Q_1 1 2 \to Q} +  P_{123 \to G}
+ P_{2 3 \overline{Q}_2 \to \overline{Q}} \right ) \,
\Big |\S^3_\mu V^\mu\Big |^2, \nonumber \\
\Big |\S_\mu(Q_1;1,2,\tilde{3};\overline{Q}_2) V^\mu\Big |^2 & \to &
\left( P_{Q_1 1 2 \to Q}
+ P_{12 \overline{Q}_2 \to \overline{Q}}
+ P_{Q_1 \tilde{1}\tilde{3} \to Q}
+ P_{\tilde{2}\tilde{3}\overline{Q}_2 \to \overline{Q}}
\right)
\, \Big |\S^3_\mu V^\mu\Big |^2, \nonumber \\
\Big |\S_\mu(Q_1;\tilde{1},\tilde{2},\tilde{3};\overline{Q}_2)
 V^\mu\Big |^2 & \to & \frac{1}{2}
\sum_{P(1,2,3)}  \left(
P_{Q_1\tilde{1}\tilde{2} \to Q}  \,
+ P_{\tilde{2}\tilde{3}\overline{Q}_2 \to \overline{Q}}
\right)
 \Big |\S^3_\mu V^\mu\Big |^2.
\end{eqnarray}
The four-quark current contributions are more complicated, but in each case
yield factors multiplying the two-quark current,
\begin{eqnarray}
&&
\Big | \T^A_\mu(Q_1,\overline{Q}_2;Q_3,\overline{Q}_4;1)V^{\mu} \Big |^2 \to
\left( P_{1\overline{Q}_4Q_3 \to G}+P_{\overline{Q}_2Q_1 1 \to G}
\right. \nonumber \\ && \hspace{5.3cm} \left.
+P_{\overline{Q}_2Q_3\overline{Q}_4 \to \overline{Q}}^{\rm non-ident.}
+P_{Q_3\overline{Q}_2Q_1 \to Q}^{\rm non-ident.}
\right) \, \Big | \S^3_\mu V^\mu \Big |^2, \nonumber \\
&&
\Big | \T^B_\mu(Q_1,\overline{Q}_4;Q_3,\overline{Q}_2;1)V^{\mu} \Big |^2 \to
\left( P_{Q_1\tilde{1}\overline{Q}_2 \to \tilde{G}}
+P_{\overline{Q}_2Q_3\overline{Q}_4 \to \overline{Q}}^{\rm non-ident.}
+P_{Q_1\overline{Q}_4Q_3 \to Q}^{\rm non-ident.}
\right) \, \Big | \S^3_\mu V^\mu \Big |^2, \nonumber \\
&&
\Big | \overline{\T}_\mu(Q_1,\overline{Q}_2;Q_3,\overline{Q}_4;1)V^{\mu} \Big
|^2 \to
\left(
P_{Q_1\tilde{1}\overline{Q}_2  \to \tilde{G}}
+P_{\overline{Q}_2Q_3\overline{Q}_4 \to \overline{Q}}^{\rm non-ident.}
+P_{Q_1\overline{Q}_4Q_3 \to Q}^{\rm non-ident.}\right. \nonumber \\ &&
\hspace{5.3cm} \left.
+P_{Q_3\tilde{1}\overline{Q}_4  \to \tilde{G}}
+P_{\overline{Q}_4Q_1\overline{Q}_2 \to \overline{Q}}^{\rm non-ident.}
+P_{Q_3\overline{Q}_2Q_1 \to Q}^{\rm non-ident.}
\right) \, \Big | \S^3_\mu V^\mu \Big |^2, \nonumber \\
&&
\Re
\left(
\overline{\T}_\mu(Q_1,\overline{Q}_2;Q_3,\overline{Q}_4;1)V^{\mu}
\right)
\left(\overline{\T}_\mu(Q_1,\overline{Q}_4;Q_3,\overline{Q}_2;1)V^{\mu}\right)
^{\dagger}
\nonumber \\ &&  \hspace{3cm}
\to - \frac{1}{2} \left(
 P_{Q_1\overline{Q}_2Q_3  \to Q}^{\rm ident.}
+P_{Q_3\overline{Q}_4Q_1  \to Q}^{\rm ident.}
+P_{\overline{Q}_2Q_3\overline{Q}_4 \to \overline{Q}}^{\rm ident.}
+P_{\overline{Q}_4Q_1\overline{Q}_2 \to \overline{Q}}^{\rm ident.}
\right)
\, \Big | \S^3_\mu V^\mu \Big |^2,  \nonumber\\
&&
\Re \left (\T^A_\mu(Q_1,\overline{Q}_2;Q_3,\overline{Q}_4;1)V^{\mu} \right)
\left (\T^B_\mu(Q_1,\overline{Q}_2;Q_3,\overline{Q}_4;1)
V^{\mu}\right)^{\dagger}
\nonumber \\ &&  \hspace{3cm}
\to - \frac{1}{2} \left(
P_{Q_1\overline{Q}_2Q_3  \to Q}^{\rm ident.}
+P_{\overline{Q}_2Q_3\overline{Q}_4 \to \overline{Q}}^{\rm ident.}
\right)
 \Big | \S^3_\mu V^\mu \Big |^2.
\end{eqnarray}
Combining these limits and eliminating the identical particle factors
where appropriate provides the triple collinear singular factor for
the five parton squared matrix elements,
\begin{eqnarray}
\lefteqn{\frac{1}{3!} \Big |\widehat{\S}^5_\mu V^\mu\Big |^2
+
\Big |\widehat{\T}^5_\mu V^\mu\Big |^2  = \left(\frac{g^2N}{2}\right)^2
\Big |\widehat{\S}^3_\mu V^\mu\Big |^2}\nonumber \\
&\times&
\Bigg[    P_{123 \to G}  +
\left(\frac{N^2-1}{N^2}\right)
\left( P_{Q_1 1 2 \to Q}
+ P_{12 \overline{Q}_2 \to \overline{Q}}
\right)
 - \left(\frac{N^2-1}{2N^4}\right)
\left(P_{Q_1 \tilde{1}\tilde{3} \to Q}
+ P_{\tilde{2}\tilde{3}\overline{Q}_2 \to \overline{Q}}
\right)
\nonumber \\
&& +\frac{n_F}{N} \left(
 P_{1\overline{Q}_4Q_3 \to G}+P_{\overline{Q}_4Q_3 1 \to G}
-\frac{1}{N^2}
 P_{Q_3\tilde{1}\overline{Q}_4 \to \tilde{G}}\right)
\nonumber \\
&& +\frac{n_F}{N}
\left( \frac{N^2-1}{N^2} \right)
\left( P_{Q_1\overline{Q}_4Q_3  \to Q}^{\rm non-ident.}
-\frac{1}{2N}
 P_{Q_1\overline{Q}_4Q_3  \to Q}^{\rm ident.}
+ P_{\overline{Q}_2Q_3\overline{Q}_4 \to \overline{Q}}^{\rm non-ident.}
-\frac{1}{2N}
  P_{\overline{Q}_2Q_3\overline{Q}_4 \to \overline{Q}}^{\rm ident.} \right)
\Bigg ].
\end{eqnarray}

\subsection{Soft/collinear factorisation}
\label{subsec:softcol}

We now examine the configurations where one gluon is soft and two
particles are collinear.
In this case, the sub-amplitudes
factorise as,
\begin{equation}
|\A(\ldots,d,\ldots,a,b,c,\ldots)|^2 \rightarrow P_{d;abc}
|\A(\ldots,d,\ldots,P,\ldots)|^2,
\end{equation}
where gluon $a$ is soft, partons $b$ and $c$ are collinear and
either colour connected or able to undergo antenna pinching.
Parton $d$ is
the adjacent colour-connected hard parton in the antenna containing the soft
gluon.
If $a$ is symmetrised over
($\tilde a$, so QED-like) then
$d$ is the quark in that colour-line; otherwise $d$ is simply the parton
adjacent to $a$.
In this limit the collinear partons form parton $P$ and carry momentum
fractions,
\begin{equation}
p_b=xp_P, \qquad p_c=(1-x)p_P,
\end{equation}
and we write the soft/collinear factor as,
\begin{equation}
P_{d;abc}(x,s_{ab},s_{bc},s_{abc},s_{ad},s_{bd},s_{cd}).
\end{equation}
To determine the limiting behaviour,   we again consider the explicit forms of
the
$\gamma^* \to $ four and five parton squared matrix elements.
All terms that possess three of the `small' denominators $s_{ab}$, $s_{ad}$,
$s_{bc}$ and $s_{abc}$ contribute in the soft/collinear limit.
Similar results could be derived using helicity methods or by examining the
on-shell limits of the recursive gluonic and quark currents of
ref.~\cite{current}.
Alternatively, these limits can be obtained directly from the triple
collinear limits of sect.~5.1, by keeping only terms proportional to $1/w$ and
subsequently
replacing $1/w$ by $(s_{bd}+s_{cd})/s_{ad}$, $1/(1-w)$ by $1$ and $y$ by $1-x$.

In fact, in this limit we find a universal soft factor
multiplied by a collinear splitting function,
\begin{equation}
P_{d;abc}(x,s_{ab},s_{bc},s_{abc},s_{ad},s_{bd},s_{cd})
 =  S_{d;abc}(x,s_{ab},s_{bc},s_{abc},s_{ad},s_{bd},s_{cd})
 P_{bc \rightarrow P}(x,s_{bc}),
\end{equation}
where,
\begin{equation}
\label{eq:softfactor}
S_{d;abc}(z,s_{ab},s_{bc},s_{abc},s_{ad},s_{bd},s_{cd}) =
\frac{2(s_{bd}+s_{cd})}{s_{ab}s_{ad}}
\left( z + \frac{(s_{ab}+z s_{bc})}{s_{abc}} \right).
\end{equation}

A similar result holds for gluon $c$ becoming soft,
\begin{equation}
|\A(\ldots,a,b,c,\ldots,e,\ldots)|^2 \rightarrow P_{abc;e}
|\A(\ldots,P,\ldots,e,\ldots)|^2,
\end{equation}
where,
\begin{equation}
P_{abc;e} = P_{d;abc} ( a \leftrightarrow c, d \leftrightarrow e).
\end{equation}
In the case
that $b$ is soft the matrix elements do not possess sufficient singularities
since the collinear
particles $a$ and $c$ are not directly colour-connected,
\begin{equation}
|\A(\ldots,a,b,c,\ldots)|^2 \rightarrow  {\rm less~singular}.
\end{equation}
Here, there may be two powers of the small invariants in the denominator, but,
when integrated over the appropriate (small) region of phase space this
yields a vanishing contribution.
On the other hand, for QED-like subamplitudes where the gluons are symmetrised
there is a non-zero contribution when either gluon is soft.
Note that in all cases where the soft particle is a quark or antiquark,
there is also no singular contribution.

\subsubsection{Soft/collinear limit of $e^+e^- \to 5$~partons}

For the specific case of the two-quark currents,
the sum over all soft/collinear limits is easily obtained,
\begin{eqnarray}
\Big |\S_\mu(Q_1;1,2,3;\overline{Q}_2) V^\mu\Big |^2 & \to &
\left( P_{Q_1 1 2;3} +  P_{Q_1;123}+  P_{123;\overline{Q}_2}
+ P_{1; 23\overline{Q}_2} \right ) \, \Big |\S^3_\mu V^\mu\Big |^2, \nonumber
\\
\Big |\S_\mu(Q_1;1,2,\tilde{3};\overline{Q}_2) V^\mu\Big |^2 & \to &
\left(
  P_{Q_1; 1 2 \overline{Q}_2}
+ P_{1 Q_1 \tilde{3}; \overline{Q}_2}
+ P_{\tilde{3}Q_1 1; 2} \right.\nonumber \\
&&
\left.+ P_{Q_1; \tilde{3} \overline{Q}_2 2}
+ P_{Q_1 1 2; \overline{Q}_2}
+ P_{\tilde{3} \overline{Q}_2 2; 1}
\right)
\, \Big |\S^3_\mu V^\mu\Big |^2, \nonumber \\
\Big |\S_\mu(Q_1;\tilde{1},\tilde{2},\tilde{3};\overline{Q}_2)
 V^\mu\Big |^2 & \to &
\sum_{P(1,2,3)}  \left(
 P_{\tilde{1}Q_1\tilde{2}; \overline{Q}_2}
+ P_{Q_1; \tilde{2}\overline{Q}_2\tilde{3}}
\right)
\, \Big |\S^3_\mu V^\mu\Big |^2.
\end{eqnarray}
The only non-vanishing contributions in the soft/collinear limit
from the four-quark currents are,
\begin{eqnarray}
&&
\Big | \T^A_\mu(Q_1,\overline{Q}_2;Q_3,\overline{Q}_4;1)V^{\mu} \Big |^2 \to
\left( P_{Q_1; 1 \overline{Q}_4Q_3}
+P_{\overline{Q}_2Q_1 1; \overline{Q}_4}
\right) \, \Big | \S^3_\mu V^\mu \Big |^2, \nonumber \\
&&
\Big | \overline{\T}_\mu(Q_1,\overline{Q}_2;Q_3,\overline{Q}_4;1)V^{\mu} \Big
|^2 \to \left(
 S_{Q_1 1 \overline{Q}_2}P_{Q_3\overline{Q}_4 \to G}
+ S_{Q_3 1 \overline{Q}_4}P_{Q_1\overline{Q}_2 \to G} \right.
\nonumber \\ && \qquad\qquad
+P_{Q_1; 1 \overline{Q}_4 Q_3}
-P_{Q_1; 1 Q_3 \overline{Q}_4}
-P_{\overline{Q}_2; 1 \overline{Q}_4 Q_3}
+P_{\overline{Q}_2; 1 Q_3 \overline{Q}_4}
\nonumber \\ && \left. \qquad\qquad
+P_{Q_3; 1 \overline{Q}_2 Q_1}
-P_{Q_3; 1 Q_1 \overline{Q}_2}
-P_{\overline{Q}_4; 1 \overline{Q}_2 Q_1}
+P_{\overline{Q}_4; 1 Q_1 \overline{Q}_2}
\right) \, \Big | \S^3_\mu V^\mu \Big |^2,
\end{eqnarray}
where the second term also includes both the unconnected soft/collinear
contribution and interferences amongst
the various subamplitudes.  This
is akin to the case of single soft gluon emission where we have,
$$
\A(\ldots,a,g,b,\ldots) \to
\left(\frac{p_a\cdot\epsilon}{p_a \cdot p_g}-\frac{p_b\cdot\epsilon}{p_b \cdot
p_g}\right)
\A(\ldots,a,b,\ldots),
$$
and therefore,
\begin{eqnarray*}
\lefteqn{\Re \left(\A(\ldots,a,g,b,\ldots,c,d,\ldots)\right)
\left(\A (\ldots,a,b,\ldots,c,g,d,\ldots)\right)^\dagger}
\\
&&
\qquad\to \frac{1}{2} \left(
S_{agc}-S_{agd}-S_{bgc}+S_{bgd}
\right)
|\A(\ldots,a,b,\ldots,c,d,\ldots)|^2.
\end{eqnarray*}
Here,  the soft factors are generated by the interference of the two eikonal
factors.
$$
\sum_{{\rm pols}}
\left(\frac{p_a\cdot\epsilon}{p_a \cdot p_g}-\frac{p_b\cdot\epsilon}{p_b \cdot
p_g}\right)
 \left(\frac{p_c\cdot\epsilon^*}{p_c \cdot p_g}-\frac{p_d\cdot\epsilon^*}{p_d
\cdot p_g}\right).
$$

Adding up these limits for the five parton squared matrix elements gives,
\begin{eqnarray}
\lefteqn{\frac{1}{3!} \Big |\widehat{\S}^5_\mu V^\mu\Big |^2
+
\Big |\widehat{\T}^5_\mu V^\mu\Big |^2  = \left(\frac{g^2N}{2}\right)^2
\Big |\widehat{\S}^3_\mu V^\mu\Big |^2}\nonumber \\
&\times&
\Bigg[
P_{Q_1 1 2;3} +  P_{Q_1;123}+  P_{123;\overline{Q}_2}
+ P_{1; 23\overline{Q}_2}
\nonumber \\
&&- \frac{1}{N^2}
\left(
  P_{Q_1; 12\overline{Q}_2} + P_{Q_1 12 ;\overline{Q}_2}
+ P_{\tilde{3}Q_1 1; 2} + P_{\tilde{3} \overline{Q}_2 2; 1}
\right)
 + \frac{1}{N^4}
\left(
  P_{\tilde{1}Q_1\tilde{2}; \overline{Q}_2}
+ P_{Q_1; \tilde{2}\overline{Q}_2\tilde{3}}
\right)
\nonumber \\
&&+ \frac{2n_F}{N}
P_{Q_1; 1 \overline{Q}_4Q_3}
\nonumber \\
&&- \frac{2n_F}{N^3}
\left(
S_{Q_1 1 \overline{Q}_2}P_{Q_3\overline{Q}_4 \to G}
+P_{Q_1; 1 \overline{Q}_4 Q_3}
-P_{Q_1; 1 Q_3 \overline{Q}_4}
-P_{\overline{Q}_2; 1 \overline{Q}_4 Q_3}
+P_{\overline{Q}_2; 1 Q_3 \overline{Q}_4}
\right)
\Bigg ].
\end{eqnarray}

\subsection{Two soft gluons}
\label{subsec:doublesoft}

Finally, we consider the contributions where two colour connected gluons
are simultaneously soft.
This was first studied by Berends and Giele \cite{multsoft} and we
include this contribution here for the sake of completeness.
Similar results have been discussed by Catani \cite{catani}.
For gluons $b$ and $c$ soft the colour ordered subamplitudes factorise,
\begin{equation}
|\A(\ldots,a,b,c,d,\ldots,)|^2  \to
 S_{abcd}(s_{ad},s_{ab},s_{cd},s_{bc},s_{abc},s_{bcd})\,
|\A(\ldots,a,d,\ldots)|^2.
\end{equation}
where the connected double soft gluon function is given by,
\begin{eqnarray}
\lefteqn{
S_{abcd}(s_{ad},s_{ab},s_{cd},s_{bc},s_{abc},s_{bcd})=
\frac{8s_{ad}^2}{s_{ab}s_{bcd}s_{abc}s_{cd}}}\nonumber \\
&&+ \frac{8s_{ad}}{s_{bc}} \left( \frac{1}{s_{ab}s_{cd}}
 + \frac{1}{s_{ab}s_{bcd}} + \frac{1}{s_{cd}s_{abc}}
 - \frac{4}{s_{abc}s_{bcd}} \right)
+ \frac{8(1-\e)}{s_{bc}^2} \left( \frac{s_{ab}}{s_{abc}}
 + \frac{s_{cd}}{s_{bcd}} -1 \right)^2.
\end{eqnarray}
Here $a$ and $d$ are the hard partons surrounding the soft pair and
may either be gluons or quark/antiquarks.
In four dimensions, the double soft factor can be extracted from
\cite{multsoft} by
squaring and summing the helicity amplitudes for two adjacent soft gluons.
Alternatively, it can be obtained by explicitly taking the double soft limit
of squared matrix elements for processes involving more than two gluons.

\subsubsection{Double soft limit of $e^+e^- \to 5$~partons}

As before, for the specific case of the two-quark currents,
the connected double soft limit is easily obtained and
summing over all contributions we find,
\begin{eqnarray}
\Big |\S_\mu(Q_1;1,2,3;\overline{Q}_2) V^\mu\Big |^2 & \to &
\left( S_{Q_1 123} + S_{123\overline{Q}_2} \right )
\, \Big |\S^3_\mu V^\mu\Big |^2, \nonumber \\
\Big |\S_\mu(Q_1;1,2,\tilde{3};\overline{Q}_2) V^\mu\Big |^2 & \to &
 S_{Q_1 1 2 \overline{Q}_2}
\, \Big |\S^3_\mu V^\mu\Big |^2, \nonumber \\
\Big |\S_\mu(Q_1;\tilde{1},\tilde{2},\tilde{3};\overline{Q}_2)
 V^\mu\Big |^2 & \to & {\rm less~singular}.
\end{eqnarray}
Combining these limits yields the double soft
singular factor for the full squared matrix elements,
\begin{equation}
 \frac{1}{3!} \Big |\widehat{\S}^5_\mu V^\mu\Big |^2
+
\Big |\widehat{\T}^5_\mu V^\mu\Big |^2  = \left(\frac{g^2N}{2}\right)^2
\Big |\widehat{\S}^3_\mu V^\mu\Big |^2
{}~\left[      S_{Q_1 123} + S_{123\overline{Q}_2}
  - \frac{1}{N^2} S_{Q_1 1 2 \overline{Q}_2}
\right ].
\end{equation}

\section{Summary}
\setcounter{equation}{0}
\label{sec:conc}

In this paper we have examined the factorisation properties of squared tree
level matrix
elements when two particles are unresolved within the framework of QCD.
At next-to-next-to-leading order,
this knowledge is required for the analytic isolation of infrared singularities
of $n+2$ parton scattering processes and subsequent numerical combination with
the one-loop $n+1$ parton (single unresolved particle) and two-loop $n$ parton
contributions.

The unresolved particles may be either soft gluons or groups
of collinear particles or combinations of both. There are four double
unresolved
cases; two soft
gluons, three simultaneously collinear particles, two independent pairs of
collinear particles
and one soft gluon together with a collinear pair.
In section~\ref{sec:unconnected} we reviewed the (trivial) factorisation that
occurs when the unresolved
particles are colour ``unconnected''. Such factorisation is well known and
involves only the familar eikonal and
Altarelli-Parisi splitting kernels used to describe single unresolved emission
(see sect.~\ref{sec:single}).

When the unresolved particles are all colour ``connected'', we find a
similar factorisation. In section~\ref{sec:connected} we introduced new
functions
to describe the triple
collinear and soft/collinear limits in addition to recalling
the known double soft
gluon limits of Berends and Giele \cite{multsoft}.
These functions are universal and apply to general multiparton scattering
amplitudes.
As a check on our results, we find that
the triple collinear splitting functions obey an expected $N=1$ SUSY identity.
In addition, in the strong ordered limit, where one particle is much more
unresolved than the other, these factors become simple products of single
unresolved factors, one associated with each unresolved particle.

To illustrate the use of these double unresolved approximations, we have
examined the singular limits of the tree level matrix elements for
$e^+e^- \to 5$~partons \cite{5jet}.   In each case, we find that in the
singular limit, the matrix elements can be approximated by a singular
factor multiplying
the tree level $e^+e^- \to 3$~parton matrix elements.
These approximations will be of use in evaluating the ${\cal O}(\alpha_s^3)$
corrections to the three jet rate in electron positron annihilation.
To achieve this however, much work still remains to be carried out.
One important ingredient is to analytically
integrate the approximations
over the unresolved regions of phase space.   A first step in this direction
has been carried out in ref.~\cite{Aude} where the hybrid subtraction method of
\cite{hybrid} has been used to evaluate the double unresolved singular
contributions
associated with a photon-gluon-quark cluster.
There is in principle no reason why this approach should not be extended to
the more general cases discussed here.
A further ingredient is the evaluation of the two loop $e^+e^- \to 3$~parton
matrix elements.   This is a formidable task in its own right and
requires analytic expressions for the two loop box graph with massless
internal and external legs for arbitrary dimension.
So far, no such expression has been found.

\section*{Acknowledgements}

We thank the CERN Theory Division for its hospitality while part
of this work was carried out and the Fermilab Theory Group
for its hospitality during the final stages.
JMC thanks the UK Particle Physics and Astronomy Research Council
for the award of a research studentship.
EWNG thanks Michael Peskin for enlightening discussions on the origins
of the $N=1$ SUSY identity.
We thank David Kosower and Stefano Catani for comments on the manuscript.


\begin{thebibliography}{99}

\bibitem{gluon}
See for example, P. S\"oding, B. Wiik, G. Wolf and S. L. Wu,
Proc. Int. Europhysics Conf. on High Energy Physics,
Brussels 1995, p 3 (1996) World Scientific
\bibitem{casimir}
B. Adeva et al, L3 Collaboration, Phys. Lett. {\bf B248} (1990) 227;\newline
P. Abreu et al, DELPHI Collaboration, Z. Phys. {\bf C59} (1993) 357;\newline
R. Akers et al, OPAL Collaboration, Z. Phys. {\bf C65} (1995) 367;\newline
R. Barate et al, ALEPH Colaboration, CERN preprint CERN-PPE/97-002.
\bibitem{egr}
J. Ellis, M.K. Gaillard and G.G. Ross,
Nucl. Phys. {\bf B111} (1976) 253.
\bibitem{ali}
A. Ali, J.G. K\"orner, G. Kramer, Z. Kunszt, G. Schierholz,
              E. Pietarinen and J. Willrodt,
 Phys. Lett. {\bf B82} (1979) 285; Nucl. Phys. {\bf B178} (1979) 421.
\bibitem{ert}
R.~K.~Ellis, D.~A.~Ross and A.~E.~Terrano,
Nucl. Phys. {\bf B178} (1981) 421.
\bibitem{fsks}
K. Fabricius, I. Schmitt, G. Kramer and G. Schierholz,
Phys. Lett. {\bf B97} (1980) 431; Z. Phys. {\bf C 11} (1981) 315.
\bibitem{5jet}
K.~Hagiwara and D.~Zeppenfeld,
Nucl. Phys. {\bf B313} (1989) 560;\\
F.~A.~Berends, W.~T.~Giele and H.~Kuijf,
Nucl. Phys. {\bf B321} (1989) 39;\\
N.~K.~Falk, D.~Graudenz and G.~Kramer,
Nucl. Phys. {\bf B328} (1989) 317.
\bibitem{KN}
P. Nason and Z. Kunszt,
in `Z Physics at LEP1', CERN Yellow report CERN 89-08 (1989).
\bibitem{resum}
S. Catani, G. Turnock and B.R. Webber,
Phys. Lett. {\bf B295} (1992) 269;\\
S. Catani, L. Trentadue, G. Turnock and  B.R. Webber,
Nucl. Phys. {\bf B407} (1993) 3.
\bibitem{schmelling}
See for example, M. Schmelling, Proc. 28th International Conference on
High Energy Physics, Warsaw, Poland, July 1996, Eds. Z. Ajduk and
A.K. Wroblewski, World Scientific 1997, p91.
\bibitem{siggi}
P.A. Movilla Fern\'andez, O. Biebel, S. Bethke, S. Kluth,
P. Pfeifenschneider and the JADE Collaboration,
`A Study of Event Shapes and Determinations of $\alpha_s$ using data
of $e^+e^-$ Annihilations at $\sqrt{s} = 22$ to 44 GeV',
preprint hep-ex/970834.
\bibitem{opal}
OPAL Collaboration, K. Ackerstaff  et al., `QCD studies with $e^+e^-$
annihilation data at 172~GeV', Contribution to XVIII International
Symposium on Lepton-Photon Interactions, Hamburg, Germany, July 1997,
OPAL-PN281.
\bibitem{GM} E.~W.~N.~Glover and D.~J.~Miller,
Phys. Lett. {\bf B396} (1997) 257.
\bibitem{BDKW}
Z.~Bern, L.~Dixon,  D.~A.~Kosower and S.~Wienzierl,
Nucl.~Phys.~{\bf B489} (1994) 3.
\bibitem{CGM} J.~M.~Campbell, E.~W.~N.~Glover and D.~J.~Miller,
`The One-loop QCD Corrections for $\gamma^* \to q\bar q gg$',
preprint hep-ph/9706297.
\bibitem{BDKall}
Z.~Bern, L.~Dixon and D.~A.~Kosower,
`One-loop Amplitudes for $e^+e^-$ to Four Partons',
preprint hep-ph/9708239.
\bibitem{DS}
L.~Dixon and A.~Signer,
Phys. Rev. Lett. {\bf 78} (1997) 811;\\
A.~Signer,
`Next-to-Leading Order Corrections to $e^+ e^- \to$ Four Jets',
hep-ph/9705218;\\
L.~Dixon and A.~Signer,
`Complete ${\cal O}(\alpha_s^3)$ Results for $e^+e^- \to (\gamma,Z) \to$ Four
Jets',
preprint hep-ph/9706285.
\bibitem{NT}
Z.~Nagy and Z. Tr\'ocs\'anyi,
`Next-to-Leading Order Calculation of Four-Jet Shape Variables',
preprint hep-ph/9707309;
`Four-jet production in $e^+e^-$ annihilation at next-to-leading order',
preprint hep-ph/9708344.
\bibitem{NTcasimir}
Z.~Nagy and Z. Tr\'ocs\'anyi,
`Group independent colour decomposition of next-to-leading order
matrix elements for $e^+e^-\to$ four partons',
preprint hep-ph/9708342.
\bibitem{NTgluino}
Z.~Nagy and Z. Tr\'ocs\'anyi,
`Excluding light gluinos using four-jet LEP events: a
next-to-leading order result',
preprint hep-ph/9708343.

\bibitem{slice}
W.T.~Giele and E.W.N.~Glover, Phys. Rev. {\bf D46} (1992) 1980;\\
W.T. Giele, E.W.N. Glover and D.A. Kosower, Nucl. Phys. {\bf B403} (1993) 633 .

\bibitem{subtract}
S.~Frixione, Z.~Kunszt and A.~Signer, Nucl. Phys. {\bf B467} (1996) 399;\\
S.~Catani and M.H.~Seymour, Nucl. Phys. {\bf B485} (1997) 291;\\
Z.~Nagy and Z.~Tr\'{o}cs\'{a}nyi,  Nucl. Phys. {\bf B486} (1997) 189.

\bibitem{hybrid}
E.W.N. Glover and M.R. Sutton, Phys. Lett. {\bf B342} (1995) 375.


\bibitem{loopcol}
Z. Bern, G. Chalmers, L. Dixon, and D.A. Kosower, Phys. Rev. Letts.
{\bf 72} (1994) 2134;\\
Z. Bern, L. Dixon, D.C. Dunbar and D.A. Kosower, Nucl. Phys. {\bf B425} (1994)
217;\\
Z. Bern, L. Dixon, and D.A. Kosower, Nucl. Phys. {\bf B437} (1995) 259;\\
Z. Bern and G. Chalmers,
Nucl. Phys. {\bf B447} (1995) 465.

\bibitem{multsoft} F.A. Berends and W.T. Giele, Nucl. Phys. {\bf B313} (1989)
595.

\bibitem{coldec}
J.E. Paton and H.-M. Chan, Nucl. Phys. {\bf B10} (1969) 519;\\
P. Cvitanovic, P.G. Lauwers and P.N. Scharbach, Nucl. Phys. {\bf B186} (1981)
165;\\
F.A. Berends and W.T. Giele, Nucl. Phys. {\bf B294} (1987) 700;\\
D.A. Kosower, B.-H. Lee and V.P. Nair, Phys. Lett. {\bf B201} (1988) 85;\\
M. Mangano, S. Parke and Z. Xu, Nucl. Phys. {\bf B298} (1988) 653;\\
M. Mangano, Nucl. Phys. {\bf B309} (1988) 461;\\
D. Zeppenfeld, Int. J. Mod. Phys. {\bf A3} (1988) 2175.

\bibitem{YFS} D.R. Yennie, S.C. Frautschi and H. Suura, Annals of Physics
              {\bf 13} (1961) 379.
\bibitem{current}
F.A. Berends and W.T. Giele, Nucl. Phys. {\bf B306} (1988) 759;\\
W.T. Giele, Ph. D. Thesis, University of Leiden (1989).

\bibitem{AP}
G.~Altarelli and G.~Parisi,
Nucl. Phys. {\bf B126} (1977) 298.

\bibitem{BN}  F. Bloch and A. Nordsieck, Phys. Rev. {\bf 52} (1937) 54.

\bibitem{KLN} T. Kinoshita, J. Math. Phys. {\bf 3} (1962) 650;\\
              T.D. Lee and M. Nauenberg, Phys. Rev. {\bf 133} (1964) 1549.

\bibitem{CDR}  C.G. Bollini and J.J. Giambiagi, Phys. Lett. {\bf 40B}
              (1972) 566;\\
              J.F. Ashmore, Nuovo Cim. Lett. {\bf 4} (1972) 289;\\
J.C. Collins, {\em Renormalisation}, Cambridge University Press (1984).

\bibitem{jetcalculus} J. Kalinowski, K. Konishi and T.R. Taylor,
Nucl. Phys. {\bf B181} (1981) 221, Nucl. Phys. {\bf B181} (1981) 253.

\bibitem{NLLJET}
K. Kato and T. Munehisa,
Comput. Phys. Commun. {\bf 64} (1991) 67.


\bibitem{Knowles}
I.~G.~Knowles, Nucl. Phys. {\bf B304} (1988) 767.

\bibitem{Aude}
A.~Gehrmann-De~Ridder and E.~W.~N.~Glover,
`A complete ${\cal O}(\alpha \alpha_s)$ calculation of the
Photon +~1 Jet Rate
in $e^{+}e^{-}$ annihilation',
Durham preprint DTP/97/26, hep-ph/9707224.

\bibitem{catani}
S. Catani, Proceedings of Workshop on `New Techniques for
Calculating Higher Order QCD Corrections',
preprint ETH-TH/93-01, Zurich (1992).

\end{thebibliography}
\end{document}